\documentclass[structabstract]{aa}  

\pdfoutput=1
\usepackage{graphicx}
\usepackage{txfonts}
\usepackage{color}
\usepackage{natbib}
\graphicspath{{.}}

\begin{document}
\title{Formation and long-term evolution of 3D vortices\\ in protoplanetary discs}

   \author{H. Meheut
          \inst{1},
          R. Keppens          
          \inst{2},
			F. Casse
          \inst{3}
          \and
          W. Benz\inst{1}
          }

   \institute{Physikalisches Institut \& Center for Space and Habitability, Universit\"at Bern, 3012 Bern, Switzerland\\
              \email{meheut@space.unibe.ch}
         \and
         Centre for Plasma Astrophysics, Department of Mathematics, KU Leuven, Celestijnenlaan 200B, 
         3001 Heverlee, Belgium 
		\and
			AstroParticule et Cosmologie (APC), Université Paris Diderot, 10 rue A. Domon et L. 
			Duquet, 75205 Paris Cedex 13, France \\			
             }
	\titlerunning{3D vortices in protoplanetary discs} 
	\authorrunning{Meheut et al.}

 
  \abstract
   {In the context of planet formation, anticyclonic vortices have recently received lots of attention for the role they can play in planetesimals formation. Radial migration of intermediate size solids toward the central star may prevent their growth to larger solid grains. On the other hand, vortices can trap the dust and accelerate this growth, counteracting fast radial transport. Multiple effects have been shown to affect this scenario, such as vortex migration or decay.}
   {The aim of this paper is to study the formation of vortices by the Rossby wave instability and their long term evolution in a full three dimensional protoplanetary disc.
  }
   {We use a robust numerical scheme combined with adaptive mesh refinement in cylindrical coordinates, allowing to affordably compute long term 3D evolutions. We consider a full disc stratified both radially and vertically that is prone to formation of vortices by the Rossby wave instability.}
   {We show that the 3D Rossby vortices grow and survive over hundreds of years without migration. The localized overdensity which initiated the instability and vortex formation survives the growth of the Rossby wave instability for very long times. When the vortices are no longer sustained by the Rossby wave instability, their shape changes toward more elliptical vortices. This allows them to survive shear-driven destruction, but they may be prone to elliptical instability and slow decay.}
   {When the conditions for growing Rossby wave-related instabilities are maintained in the disc, large-scale vortices can 
   survive over very long timescales and may be able to concentrate solids.}


   \maketitle
%

\section{Introduction}

The origin of the kilometer size planetesimals is still an issue of planet formation theory.
The growth of micron size solids toward centimetre or meter blocks is predicted by coagulation models \citep{D09}, but collisions usually lead to destruction of such solids \citep{B00}. These collisions arise because the pressure supported gas moves at sub-Keplerian speed and the solid bodies experience a head wind from the gas due to the drag forces. The consequence is a loss of angular momentum and a radial drift of solids spiraling toward the central star. This can occur on a timescale as short as a hundred years for meter size blocks at one astronomical unit \citep{WEI77}. Such timescales are far too short to explain subsequent growth into larger particles that are unaffected by the head wind.\\

To overcome these difficulties, anticyclonic vortices, where the velocity streamlines rotate in the opposite direction to the Keplerian flow \citep{M90}, may be a good 
environment for planetesimal growth. They induce a drag force towards the centre of the structures, and are therefore proposed as possible nurseries for intermediate size blocks, concentrating the solids and accelerating the planetesimal formation processes \citep{BAR95,TBD96,BCP99,GL99-2,GL00,JAB04,HK10}. As mentioned by \citet{A11}, vortices can also be important for mechanisms that require an increase in the dust-to-gas ratio, such as the streaming instability \citep{JYM09}.
Furthermore, \citet{KB06} argued that anticyclonic vortices are regions with lower turbulence. Fragmentation is less likely to occur if lower
velocity fluctuations prevail.\\

One of the main difficulties with this scenario is the question of the formation of vortices. Different instabilities have been proposed for the vortex generation, such as the baroclinic instability \citep{KLA03,KLA04} in which interest has been recently revived \citep{LP10,LK11}, or potentially the magneto-rotational instability \citep{FN05}. In this paper we will study the Rossby wave instability (RWI), that may form vortices with unusual elongated shape in regions of particular interest for the long term survival of those structures.
It has been recently pointed out that the survival of the vortices on a sufficient timescale may be an issue. The elliptical instability may destroy three dimensional (3D) elliptical vortices \citep{LES09}. Moreover, \citet{PLP10} have shown that vortices can be subject to radial migration toward the disc centre in a short timescale. \\

In this paper, we investigate the generation, the 3D structure and the long term evolution of vortices formed by the RWI within a protoplanetary disc. This follows the work presented in \citet{MEH10}, revisited with a new code that allows to handle long term simulations. The numerical and physical setup is described in the next section. The results are presented in section \ref{results}, followed by a detailed discussion including comparisons with previous works.

   \begin{figure}
   \centering
   \includegraphics[trim=2.2cm 3cm 1.5cm 3cm ,clip=true,angle=90,width=8cm]{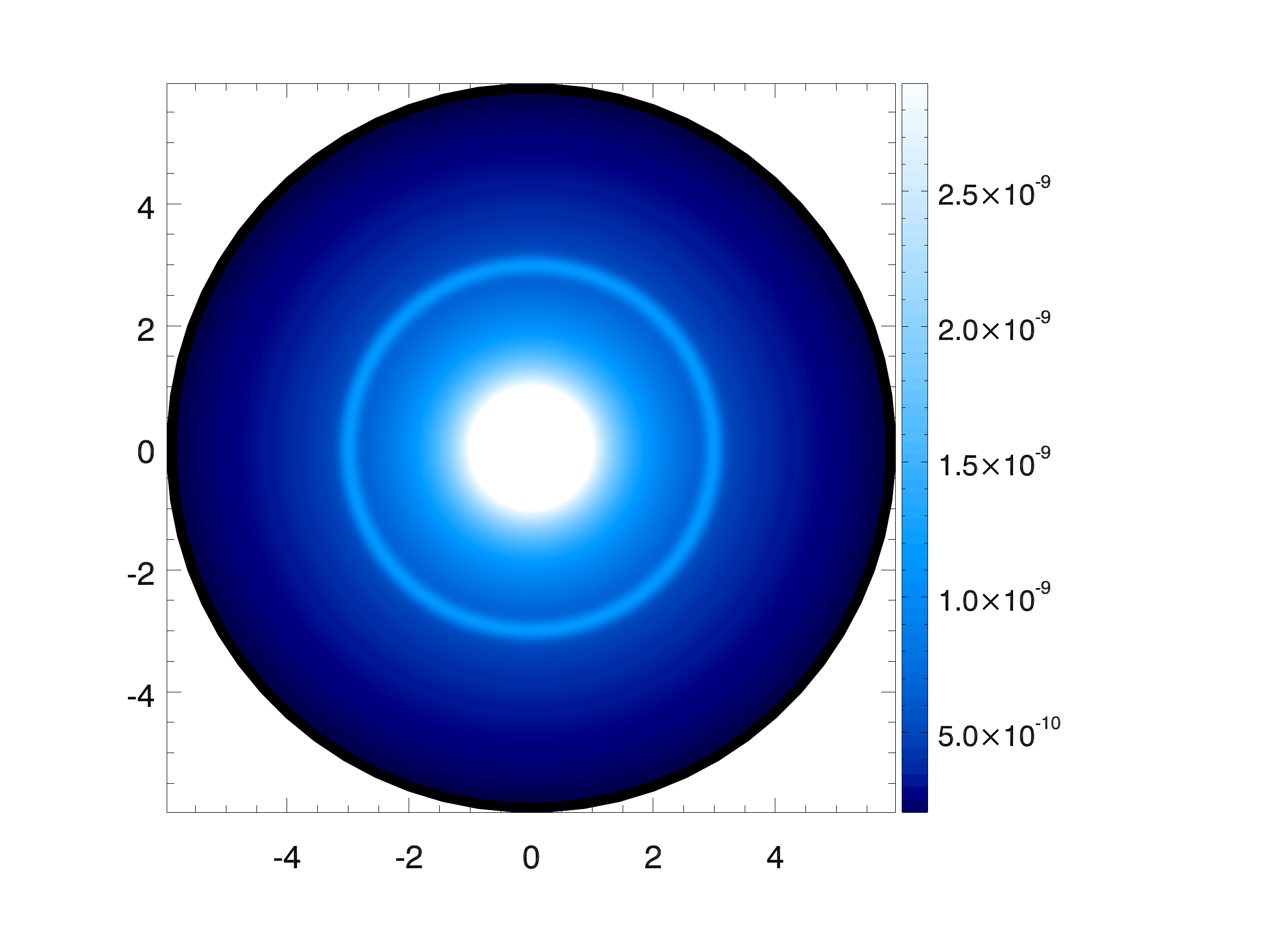}
      \caption{Mid-plane density of the gas in $g\,{\rm{cm}}^{-3}$. The density bump is placed at $3$ AU.}
         \label{Figrhomidini}
   \end{figure}


   \begin{figure*}[!]
   \centering
	\includegraphics[angle=90,trim=7.2cm 2.05cm 7.5cm 0.8cm ,clip=true,width=19cm]{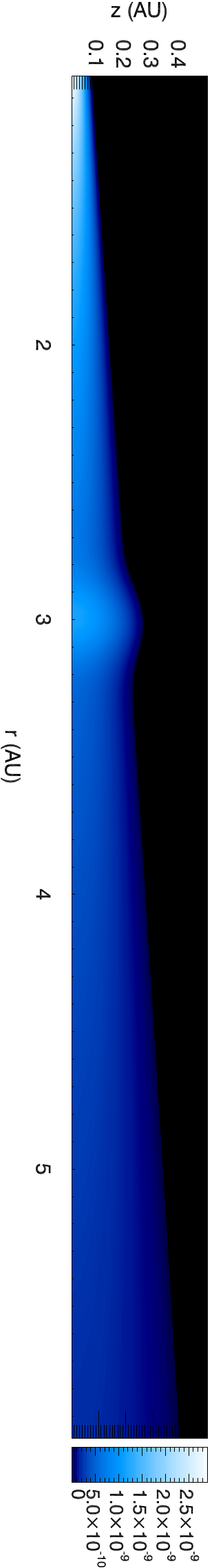}
	\includegraphics[width=16.7cm,height=1.8cm,keepaspectratio=false,trim=-0.1cm -0.2cm -0.5cm -0.4cm,clip]{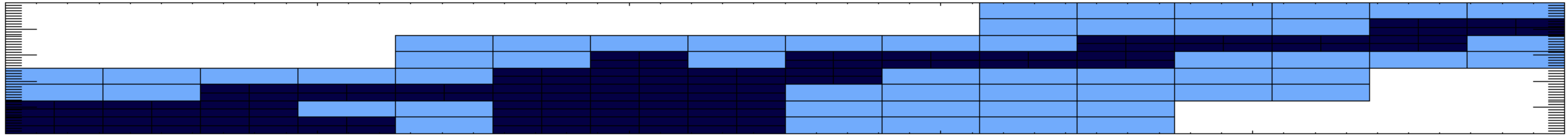}
   \caption{Initial gas density in the disc in $g\,{\rm{cm}}^{-3}$ ({\itshape up}) and AMR grid 
   (\textit{down}). The three levels are plotted in white, blue and dark blue from lower to higher 
   resolution. The highest resolution is reached in the overdensity region and at the interface 
   between the disc and corona.}
   \label{Figrhovertini}
    \end{figure*}

\section{Methods and setup}

\subsection{Rossby wave instability}

The RWI  \citep{LOV99,LI00,LI01} has been studied and discussed in various situations of differentially rotating discs, from galactic discs \citep{LOV78,SEL91} to microquasar and protoplanetary discs \citep{PAP85, LOV99}. 
It can be seen as the form that the Kelvin-Helmholtz instability takes in 
differentially rotating disks, and has a similar instability criterion: 
It requires an extremum in a vorticity related quantity, defined in a non-magnetised thin disc as \citep{LI00}:
\begin{equation}
\mathcal L= \frac{\Sigma \Omega}{\kappa^2} (p\Sigma^{-\gamma})^{2/\gamma} =	
\frac{\Sigma}{2(\vec\nabla\times\vec v)_z}(p\Sigma^{-\gamma})^{2/\gamma} \,,
\end{equation}
where $\Sigma$ is the surface density, $p$ the pressure, $\vec v$ is the velocity of the fluid, $\gamma$ the adiabatic index, $\Omega$ the rotation frequency and $\kappa^2 = 4\Omega^2 + 2r\Omega\Omega '$ the squared epicyclic frequency (so that $\kappa^2/2\Omega$ is the vorticity). Here the prime denotes a radial derivative. For the isentropic discs we consider here, this criterium is reduced to an extremum of $\frac{\Sigma \Omega}{\kappa^2}$. This quantity is directly related to vortensity
defined as the ratio of vorticity to surface density.

\subsection{Equations}

We work in cylindrical coordinates $(r, z, \varphi)$ with the 3D Euler equations
\begin{eqnarray}
\partial_t\rho+\vec \nabla\cdot (\vec v\rho)=0 \,,\\
\partial_t(\rho\vec v)+\vec\nabla\cdot(\vec v\rho\vec v)+\vec\nabla p=-\rho\vec\nabla \Phi_G \,,
\end{eqnarray}
where $\rho$ is the mass density of the fluid, $\vec v$ its velocity, and $p$ its pressure. $\Phi_G = -\frac{GM_*}{\sqrt{r^2 + z^2}}$ is the gravity potential of the central object with $G$ the gravitational constant and $M_*$ the mass of the star. We consider a barotropic flow, i.e. the entropy is constant in the entire system. The pressure is then $p = S\rho^\gamma$, with the adiabatic index $\gamma= 5/3$ and the constant $S$ related to entropy. The sound speed is given by $c_s^2=\gamma p/\rho=S\gamma \rho^{\gamma-1}$ and the temperature by $T\sim p/\rho=S\rho^{\gamma-1}$. The temperature is normalised to the temperature at $1~AU$.

\subsection{Numerics}

The numerical methods used for these simulations are inspired by those of \cite{MEH10}. We use the Message Passing Interface-Adaptive Mesh Refinement Versatile Advection Code (MPI-AMRVAC) developed by Keppens and Meliani \citep{KEP11}. The use of a code that allows the mesh to adapt during the simulation has multiple advantages. One aim is to reach higher resolution in the vortex regions, as well as to have a better resolution in the upper region of the disc, where the gas density decreases abruptly forming a `corona'. The use of AMR offers flexibility to enforce a lower resolution in this physically irrelevant, but computationally challenging, region. Since low density material can easily be accelerated, in turn enforcing smaller timesteps to maintain numerical stability in explicit time stepping schemes, handling the corona at low resolution is computationally advantageous. The current AMR approach can therefore model the unstable disc on long timescales with similar computing resources as the short timescale run done earlier. Also, by better numerically representing the disc-corona interface, combined with a sharper limiter function, we have a more robust numerical treatment of the governing equations.

The numerical scheme is the same for all refinement levels, namely the Total Variation Diminishing Lax-Friedrich scheme (see \cite{TOO96}) with a third order accurate Koren limiter \citep{KOR93} on the primitive variables. We use a cylindrical grid with $r~\epsilon~ [1, 6]$ AU, $z~\epsilon~ [0, 0.5]$ AU and the full azimuthal direction $\varphi~\epsilon ~[0, 2\pi]$. The simulations consider only the upper half of the disc, as the disc mid-plane appeared to be a symmetry plane for the instability in our earlier full vertical simulations \citep{MEH10}. If not specified, the length is given in AU and the code unit time corresponds to $1/(2\pi)$ yr.
The resolution of the base level is $(64,32,32)$. Up to three levels of refinement are allowed, corresponding to an effective resolution of $(256,128,128)$. In the region of the density bump where the instability is expected to grow, the resolution is fixed to the higher level during the whole simulation. The initial grid is presented in Fig.~\ref{Figrhovertini}, and then evolves to follow the growth of the instability. The refinement criterion being the density variation, the grid follows the growth of the spiral density waves propagating radially. At the end of the simulation $\sim 70\%$ of the grid volume is at the highest resolution level. In the beginning, it is only about $\sim 26\%$, as visually clear from Fig.~\ref{Figrhovertini}. Since refinement happens in all directions, each grid block handled at the base resolution level represents a gain factor of 64 in computational cost compared to a uniform high resolution run. This gain is even more dramatic when memory issues are considered as well. During the entire run, the corona region is maintained at this lowest resolution, dramatically impacting stability, memory and wall-clock time.

The boundary conditions are transparent at inner and outer radius, we consider a mid-plane symmetry and transparent boundary conditions are also applied at the upper boundary, whereas the azimuthal direction is periodic.

\subsection{Initial conditions}

The initial conditions for the simulations are chosen to represent a protoplanetary disc in radial and vertical equilibrium. We consider that an overdensity is formed at some radius ($r_{bump}$) of the disc. This bump could have been formed due to multiple effects.

\begin{enumerate}
\item{The presence of a dead zone in the protoplanetary disc, corresponding to a region of lower ionisation and resistivity, leads to a lower accretion rate in this region \citep{G96}. This can induce the formation of an overdensity at the edges of this region \citep{VAR06,LYR08,KLG09}, which can in turn trigger the RWI. The initial density bump used in this paper, is a Gaussian fit to the one obtained in \cite{VAR06}, when the instability started to grow.}
\item{The ice sublimation front ('snow line') can also be responsible for the formation of an overdensity \citep{KL07}. One can note that an extremum of entropy should be located in this region, which may also trigger the RWI \citep{LOV99}.}
\item{It has also been shown that the RWI grows at the edge of planet gaps \citep{KLL03,LP211}. }
\end{enumerate}
The radial positions of these regions depend on the physical characteristics of the disc (such as accretion rate and temperature) or the position of the planets. We have chosen to place the bump at a distance of $3$ AU which is a plausible region for these different effects.

The initial mid-plane density is given by
\begin{equation}
\rho(r,0,\varphi)=\rho_0r^\alpha\Big(1+\chi\exp\big(\frac{r-r_B}{\sqrt{2}\sigma}\big)^2\Big) \,,
\end{equation}
with $\rho_0=3.10^{-9}g\,{\rm{cm}}^{-3}$ the density at $r=1$ AU, and $\alpha=-1.5$ the power law index of the underlying density. This value for the alpha parameter gives a surface density varying approximately as $\Sigma\sim r^{-1/2}$ in the absence of the bump. Parameters $r_B=3$ AU, $\chi=1$, $\sigma=0.1$ AU are the radius, amplitude and width of the Gaussian bump. 

The vertical and radial force equilibria give the initial vertical structure in density and azimuthal velocity, respectively. The pressure is calculated with constant entropy, with $S=10^{-3}$. The mid-plane and vertical density profile chosen as initial condition are shown on Fig.~\ref{Figrhomidini} and Fig.~\ref{Figrhovertini}.

The azimuthal velocity is close to Keplerian where the deviation from Keplerian rotation is due to the pressure gradient. In most of the disc, except in the inner part of the bump, the gas is then sub-Keplerian.
There is no vertical velocity initially, whereas a very low radial velocity perturbation is added to the equilibrium as a seed for the instability. These are random perturbations, meaning that all the spatial frequencies in the three directions are present in the seed.

\section{Results}\label{results}

The evolution of the pressure bump in a protoplanetary disc is followed over more than $600$ years ($t_{max}=4000$). This timescale corresponds to $\sim 123$ rotations of the overdensity and $\sim 43$ rotations at the outer edge of the grid. Such a configuration allows the development of the Rossby wave instability and we present here the characteristics of the 3D flow under this instability and its non-linear evolution.


   \begin{figure*}
   \centering
	\includegraphics[angle=90,trim= 7cm 1cm 5.8cm 0.8cm ,clip=true,width=\textwidth]{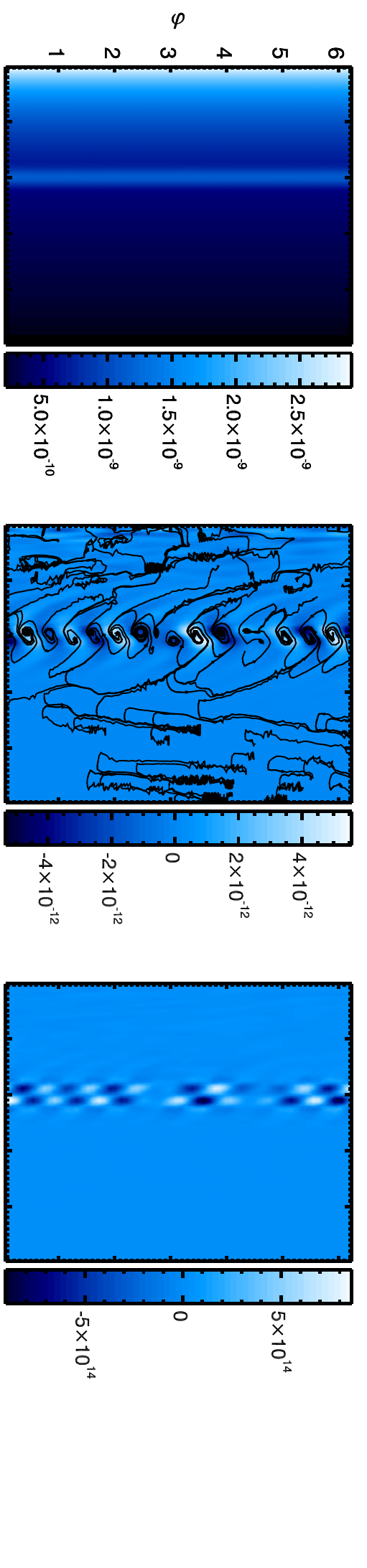}\\
	\includegraphics[angle=90,trim= 7cm 1cm 5.8cm 0.8cm ,clip=true,width=\textwidth]{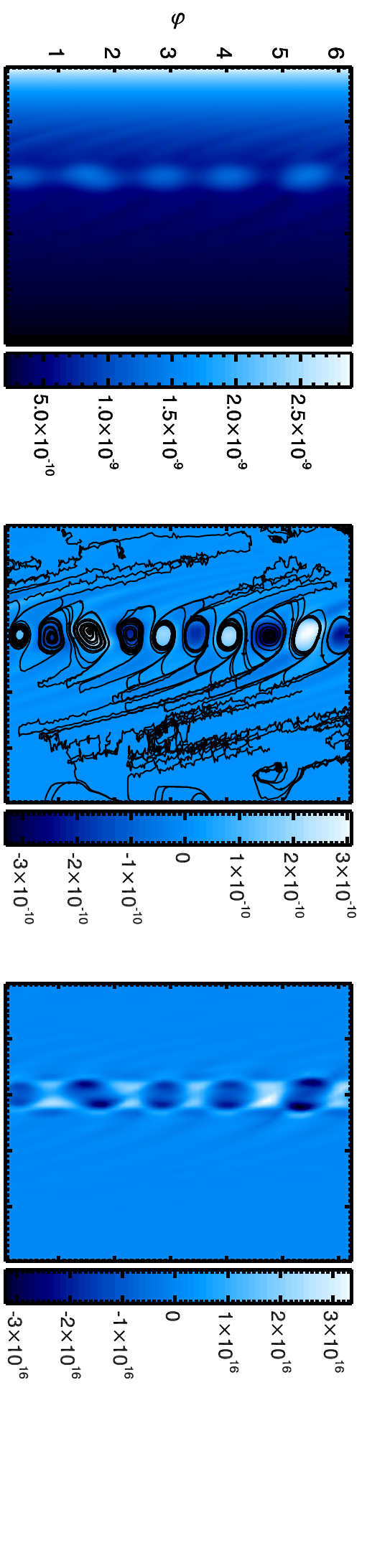}\\
	\includegraphics[angle=90,trim= 6cm 1cm 5.8cm 0.8cm ,clip=true,width=\textwidth]{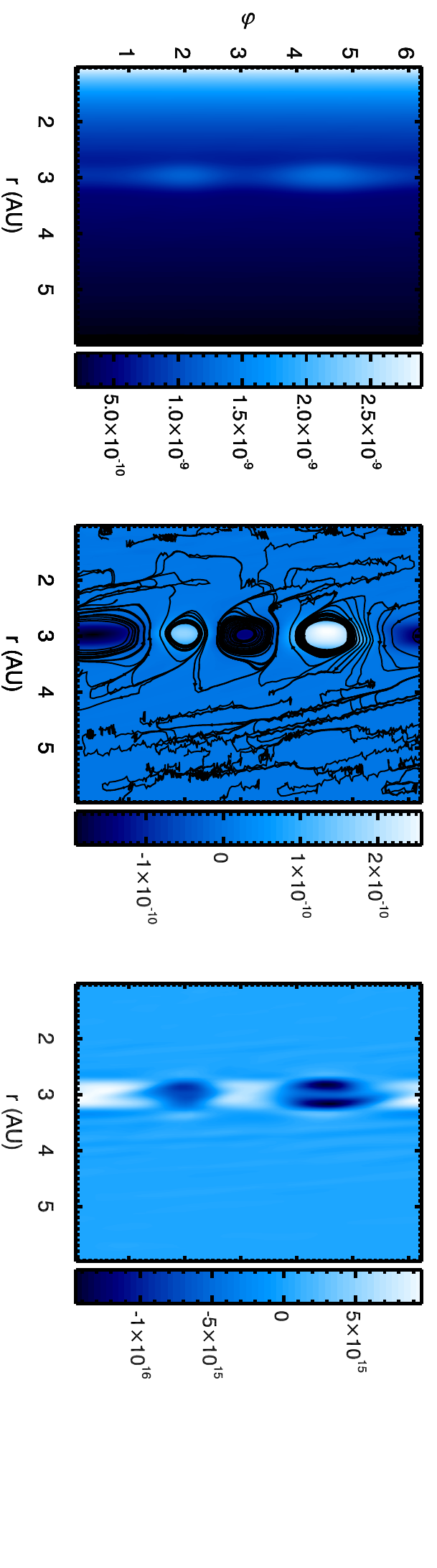}
   \caption{Density (\emph{left}), perturbation of density and velocity streamlines (\emph{center}), perturbed vortensity defined as $\zeta-\langle\zeta\rangle_\varphi$ with $\zeta=(\nabla\times v)_z/\rho$ (\emph{right}) at $t=100,300,800$ from top to bottom.}
    \label{Figevol}%
    \end{figure*}

   \begin{figure}
   \centering
   \includegraphics[width=6cm,trim=0 4cm 0 2cm,angle=90]{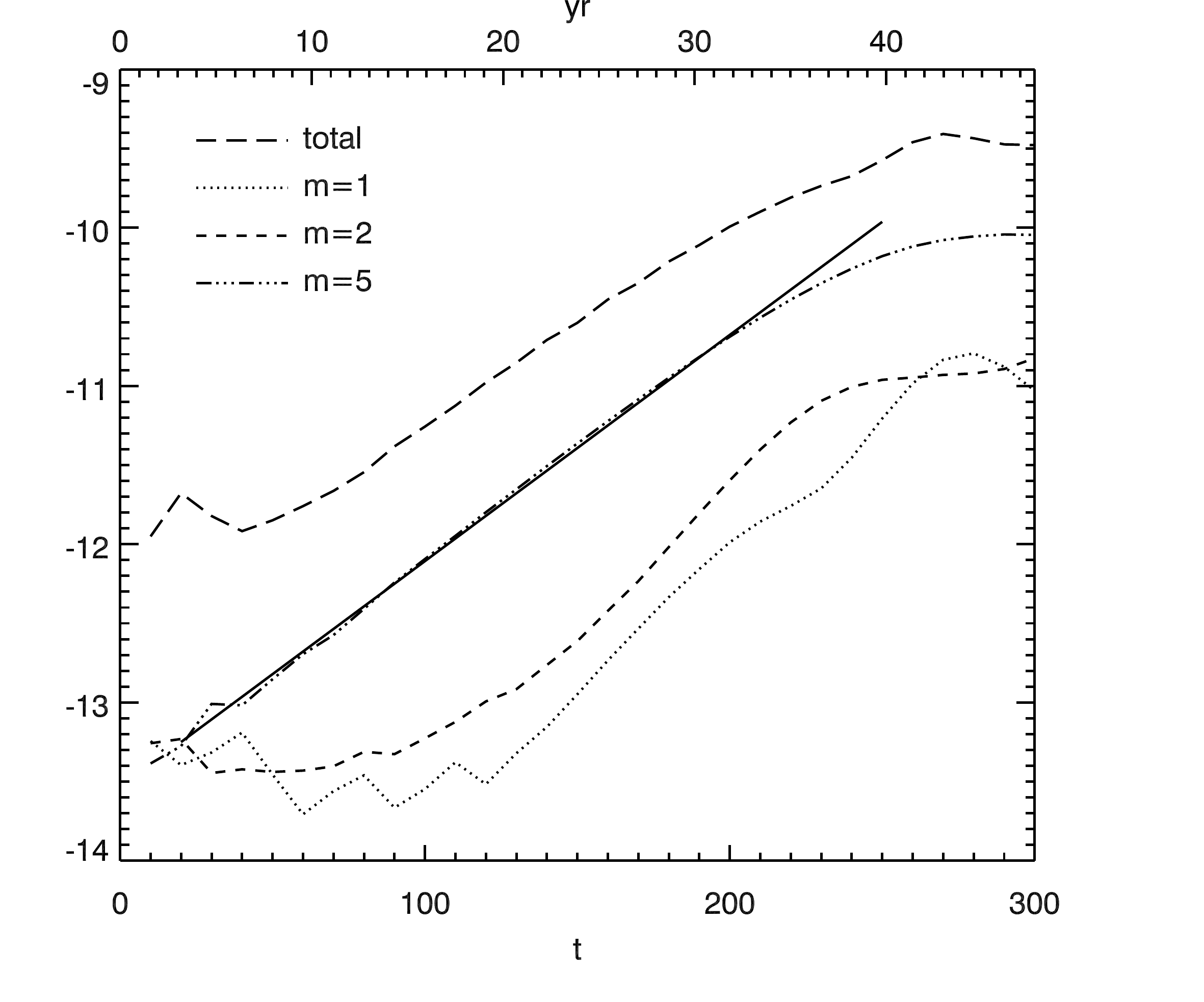}\\
   \includegraphics[width=9cm,trim=0 4cm 0 2cm,angle=90]{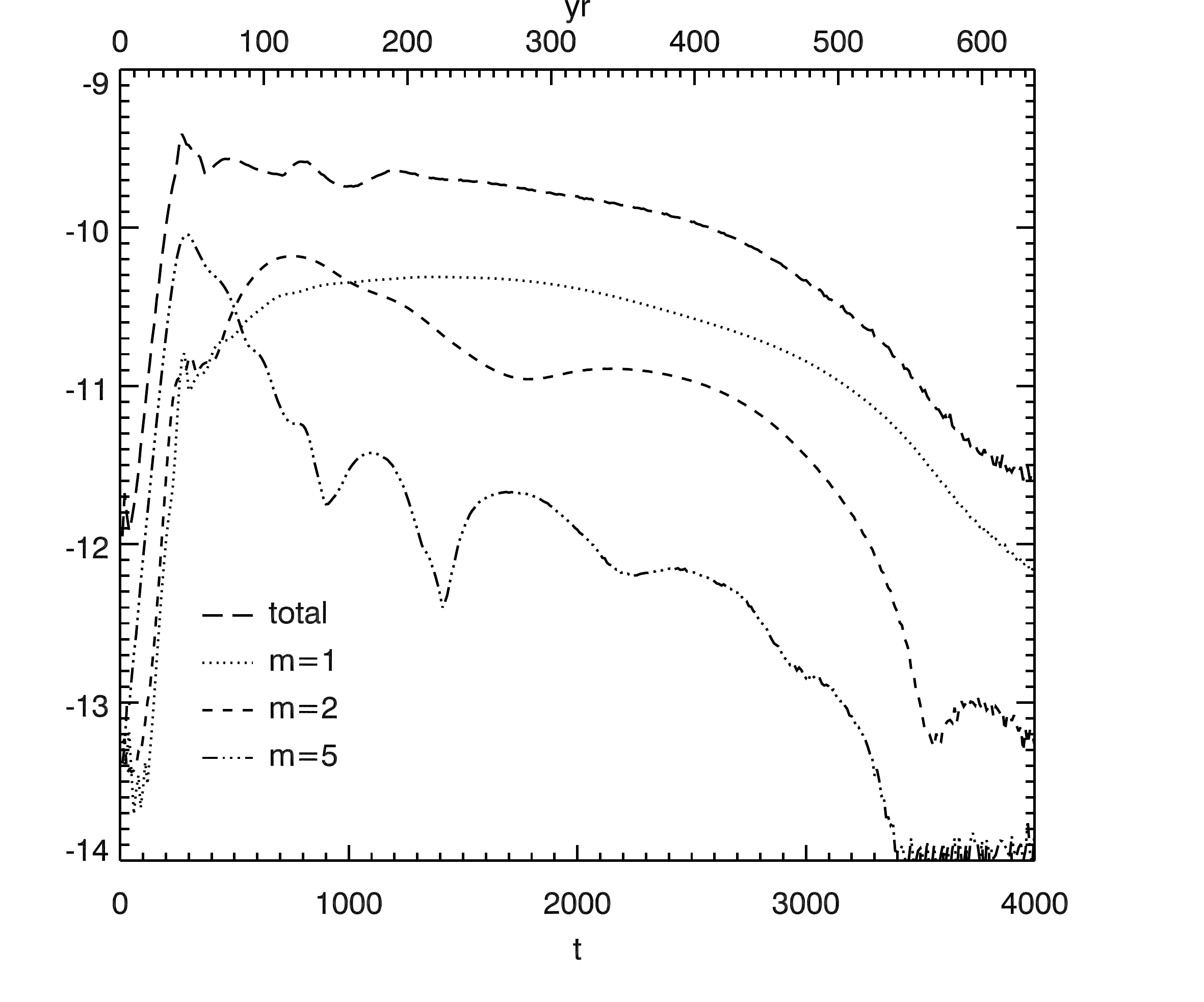}
      \caption{Time evolution of the amplitudes of the density perturbations in a 
      logarithmic scale. The time is given in code units (lower axis) and in years (upper axis).
      We also show the amplitude of some of the modes with low mode number. The upper figure 
      is a zoom on the exponential growth where we have added a linear fit for the dominant 
      mode whose growth rate is $0.033$ in code units.
              }
         \label{Figgrowth}
   \end{figure}

\subsection{Growth of the RWI}

\begin{figure}
   \centering
   \includegraphics[width=6cm,trim=0 4cm 0 2cm,angle=90]{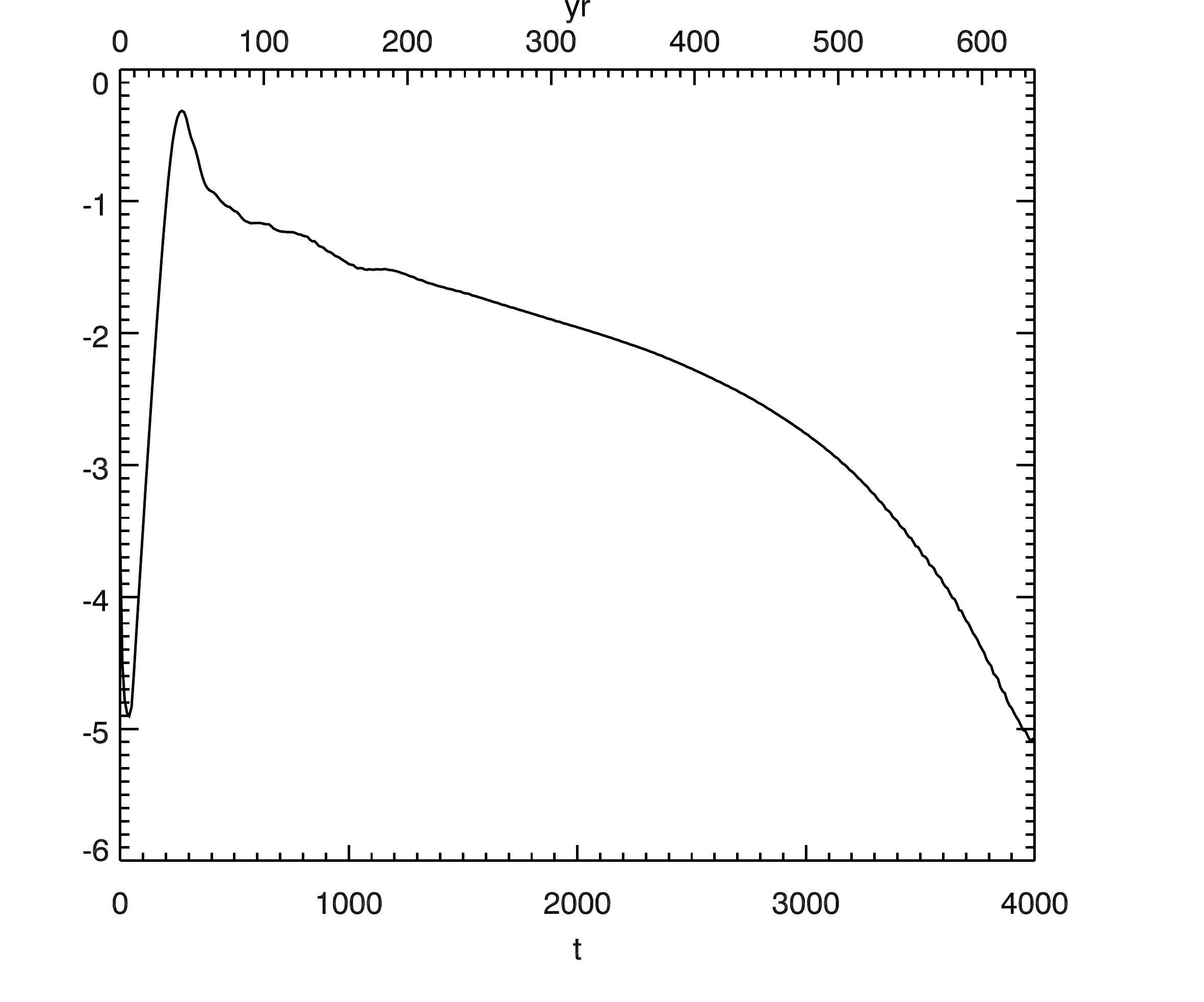}
      \caption{Time evolution of enstrophy in a 
      logarithmic scale. The time is given in code units (lower axis) and in years (upper axis).
              }
         \label{Figenstrophy}
   \end{figure}

The RWI is characterised by vortex waves in the region of the pressure bump and spiral 
density waves propagating on each side of the bump. These features can be seen on Fig.~\ref{Figevol} where the vortex waves can be identified by the vortensity extrema or the velocity streamlines characteristic for such waves. On the upper plot, the inner and outer Rossby waves are visible. Spiral density waves are emitted from each vortex inward and outward (e.g. at $t=300$). The growth of the instability is quantified in Fig.~\ref{Figgrowth} and \ref{Figenstrophy}. The linear phase of the instability corresponds to exponential growth of the perturbations until $t\sim 300$ when the density perturbation reaches  $\sim 20\%$ of the initial density. The straight line corresponds to a linear fit of the amplitude growth of the density perturbation in logarithmic scale. This fit gives a growth rate of $\gamma=0.033\Omega_0$ in code units which corresponds to $0.17\Omega(r_B)$ or $0.21\,{\rm{yr}}^{-1}$. 
We have also plotted the amplitude evolution of selected azimuthal modes, out of the Fourier transform of the density
\begin{equation}
\rho(r,z,\varphi,t)=\sum_m \rho_m(r,z,t)\exp(-im\varphi).
\end{equation}
The most unstable mode during the exponential growth has azimuthal mode number $m=5$, but lower azimuthal mode numbers are then excited by the $m=5$ mode leading to their growth, with $m=1$ and $m=2$ shown on the figure. The mode number corresponds, on Fig.~\ref{Figevol}, to the number of anticyclonic vortices, which are those counter-rotating the Keplerian rotation. They correspond to high pressure and negative vortensity regions. A high number of anticyclonic vortices come out from the initial random perturbations, rapidly decreasing to $5$. 

\subsection{Saturation of the RWI}


   \begin{figure}
   \centering
	\includegraphics[width=4.2cm,trim=0.7cm 1cm 0.8cm 0.7cm ,clip]{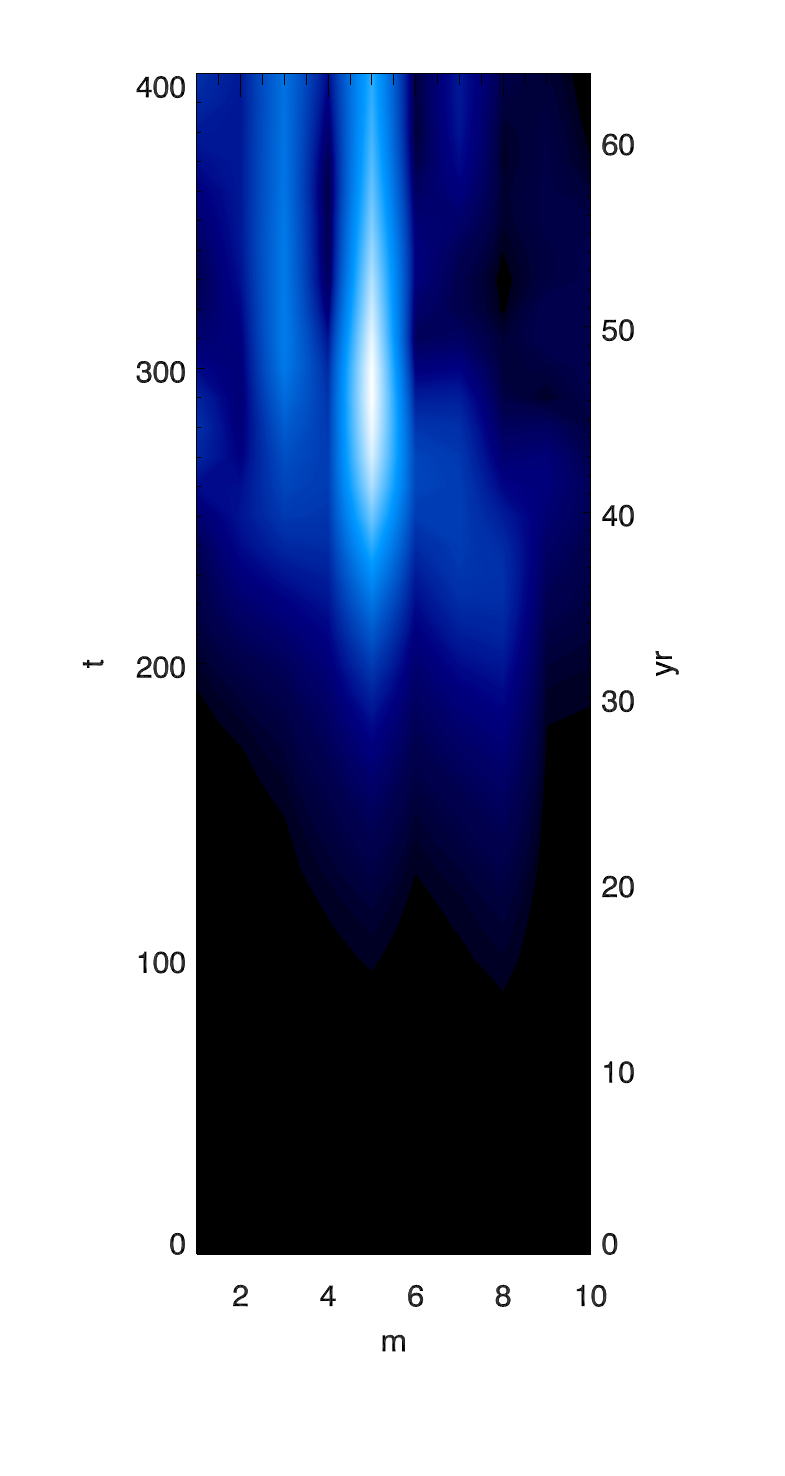}
   \includegraphics[width=4.2cm,trim=0.7cm 1cm 0.8cm 0.7cm,clip]{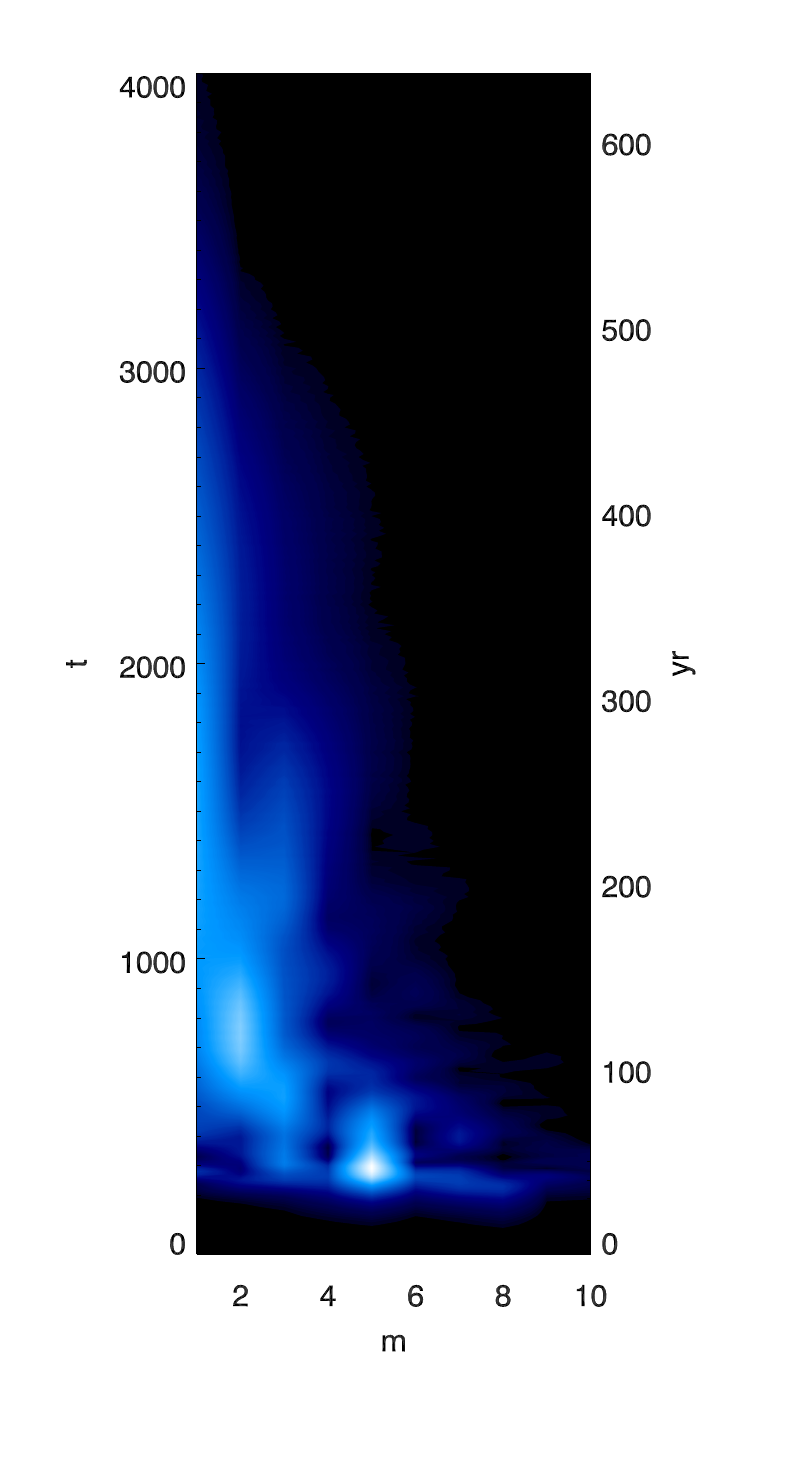}
      \caption{Comparison of the amplitude of the first $10$ azimuthal modes between $t=0$ 
      and $t=400$ (left) and over the whole simulation (right). The instability is dominated 
      by the mode $m=5$ but both lower and higher azimuthal modes are excited during 
      the growth phase. The decaying phase shows a shift to lower azimuthal mode number.
              }
         \label{Figamplmodes}
   \end{figure}

After the exponential growth, the instability reaches saturation due to non-linearities inducing mode mixing. Fig.~\ref{Figamplmodes} shows the amplitude of the modes at saturation. The fifth mode clearly dominates  for quite some time, and modes with high azimuthal mode number are less prominent, but very low and intermediate azimuthal mode numbers are non-linearly excited by the prevailing mode. However one can see on Fig.~\ref{Figevol} that later on ($t=800$) $m=2$ dominates, and then $m=1$ until the end of the simulation (see Fig.~\ref{Figamplmodes}). On the other hand, the higher mode numbers ($5<m<10$) are not dominant after saturation. 
So the general sketch of the instability is first the linear growth of the fifth mode, then saturation is reached and the larger wavelength modes (lower mode number) dominate with a transfer towards larger scales. This evolution diminishes the number of vortices through merging as described in \citet{GL99-2}. This is similar to the 2D simulations of \citet{BK03} where planar shear layers were studied in high resolution Cartesian settings. They found that the pairing/merging of vortices in Kelvin-Helmholtz unstable shear layers, studied from hydro to magnetized cases, is controlled by growth of subharmonic modes and phase differences between them. In the present 3D cylindrical configurations, their results represent local box evolutions, performed in the frame rotating at the local Keplerian speed. This pairing/merging trend to large scale structure is then predominantly a 2D process known from planar hydrodynamical evolutions. As the RWI is closely related to the Kelvin-Helmholtz instability, the merging of vortices seen in our 3D, thin disc simulations is consistent with the \citet{BK03} findings. 

   \begin{figure}
   \centering
   \includegraphics[width=6cm,trim=0cm 0 0cm 0 0,angle=90]{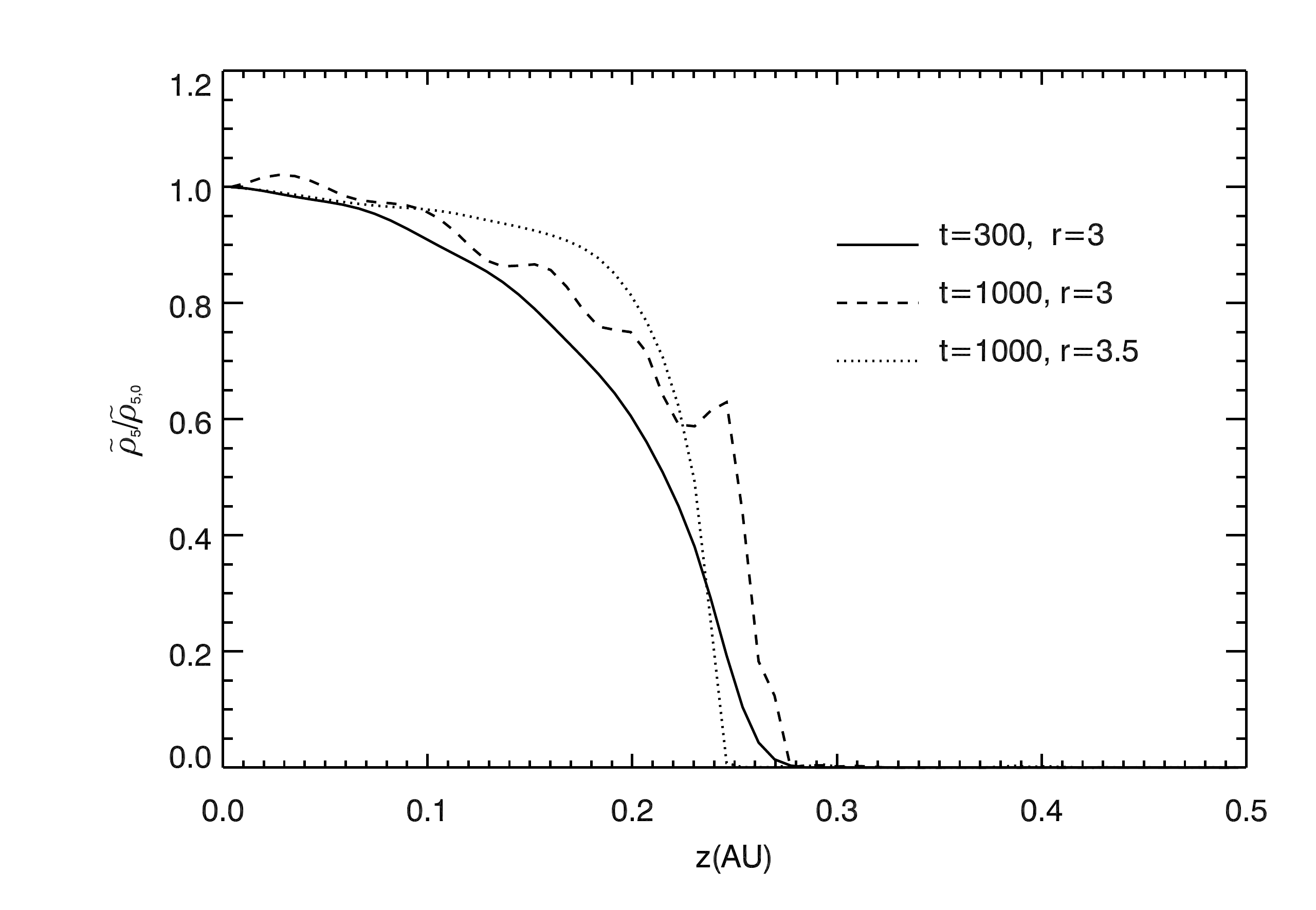}
      \caption{Vertical structure in density of the dominant fifth azimuthal mode. The amplitude of the mode is normalised to its value in $z=0$, so its structure at different times and positions can be compared.
              }
         \label{Figm5vert}
   \end{figure}

\subsection{Decay of the vortices}

The simulation has been run over more than $600$ years to investigate the long term evolution of the vortices. After the growing phase, the vortices survive during hundreds of years but then start to decrease and have disappeared at the end of the simulation. After a few hundreds of inner rotations, the dominant azimuthal mode does not change anymore, and has a global $m=1$ value. This corresponds to the largest scale possible. One can see on Fig.~\ref{Figm5vert} that during the decaying phase (after about 100-150 years), vertical mode structure appears in the $m=5$ mode. This structure is present inside the vortex ($r=3$) but not in the outer region ($r=3.5$).

   \begin{figure}
   \centering
   \includegraphics[width=6cm,trim=0cm 0 0cm 0 0,angle=90]{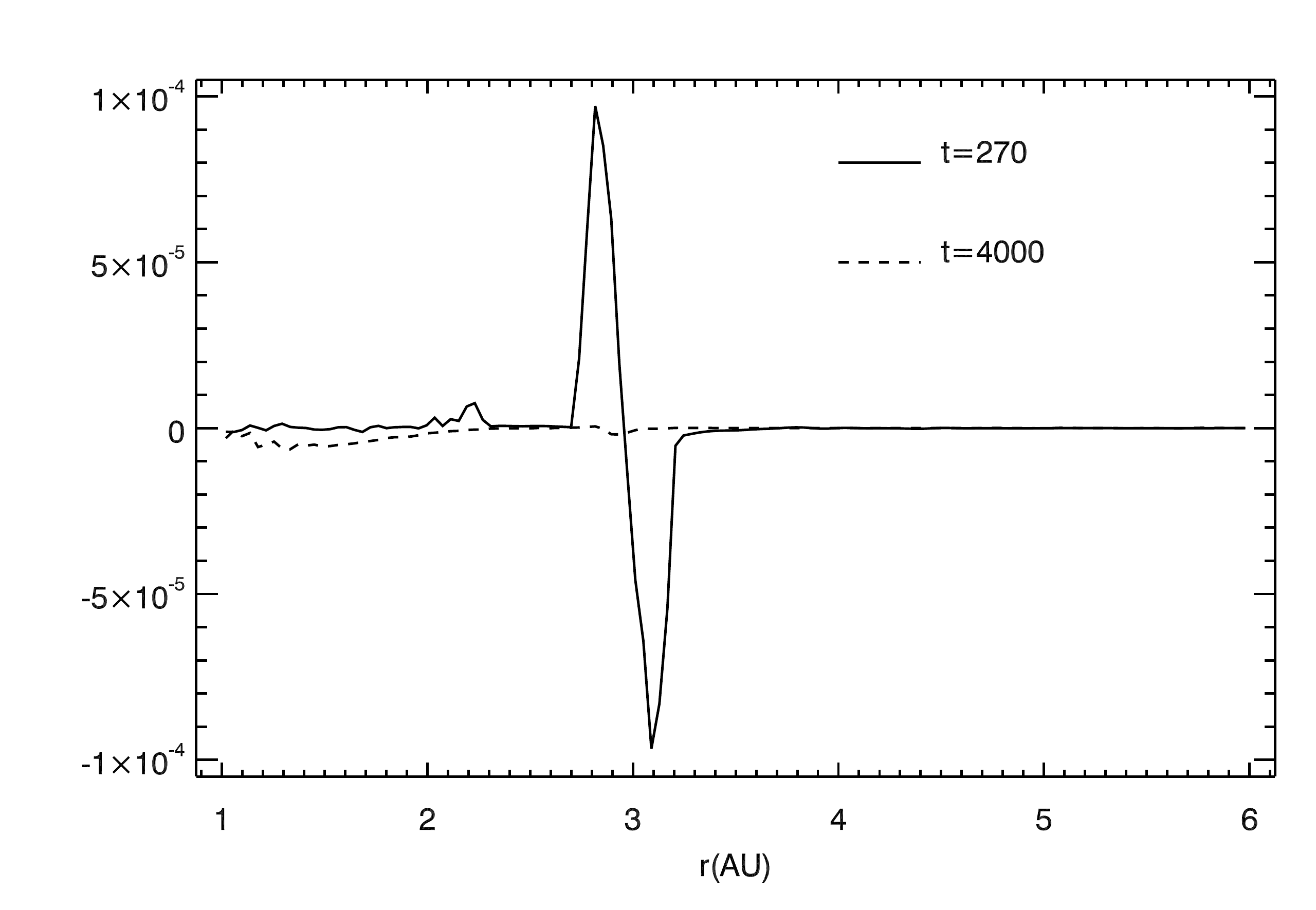}
      \caption{Accretion as a function of radius during the exponential growth (\emph{solid line}) and at the end of the simulation (\emph{dashed line}). The accretion rate is defined as $-\int\int d\varphi dz r\rho v_r$, and the amplitude is given in code units which correspond to $\sim 10^{-4}M_\odot yr^{-1}$.
              }
         \label{Figaccretion}
   \end{figure}

\begin{figure}
   \parbox{4cm}{\includegraphics[width=3.5cm,trim=0.5cm 5cm 1.5cm 2cm,angle=90,clip=true]{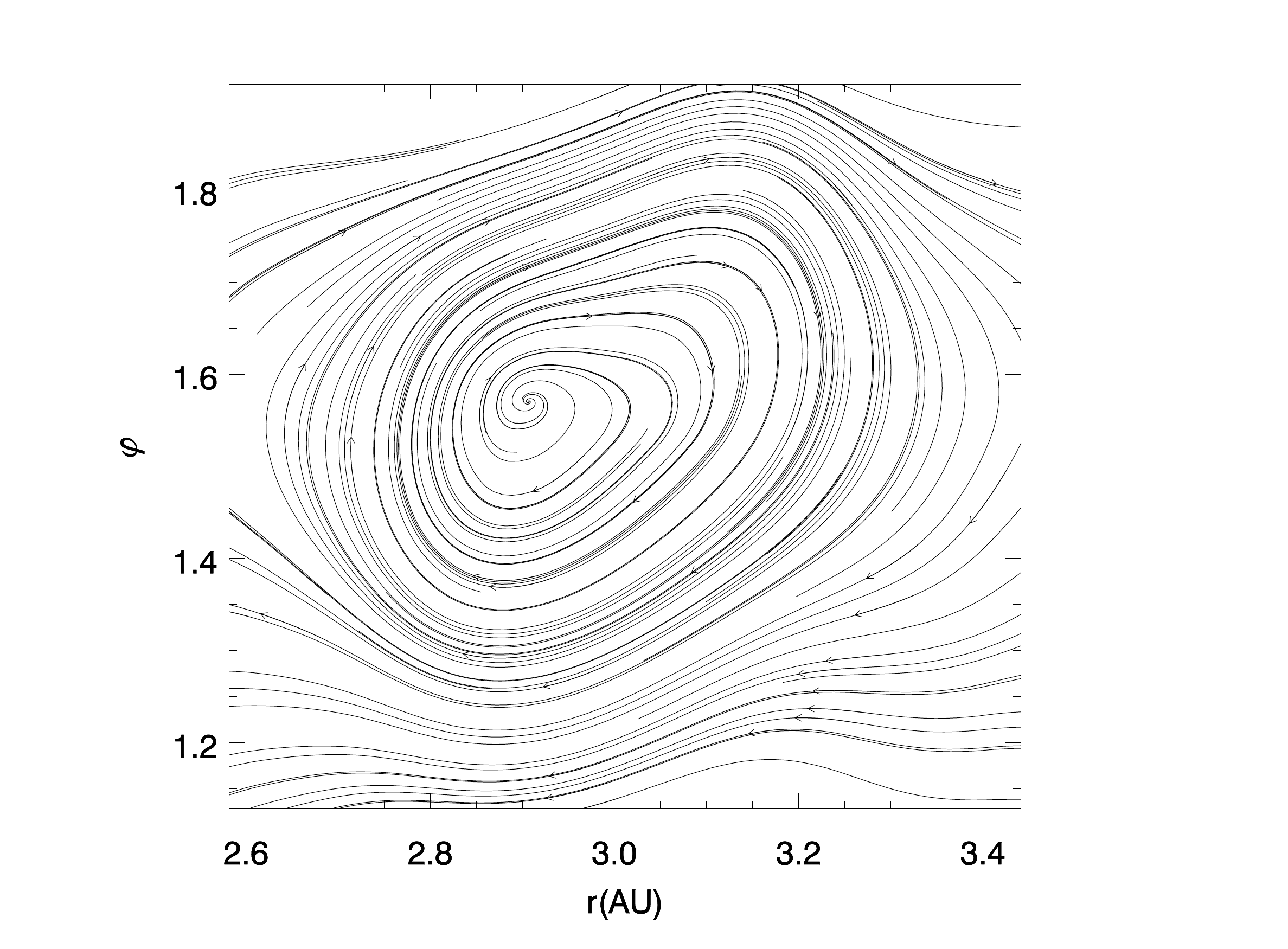}}
   \parbox{8cm}{\includegraphics[width=8cm,trim=0.5cm 9cm 1.5cm 6cm,angle=90,clip=true]{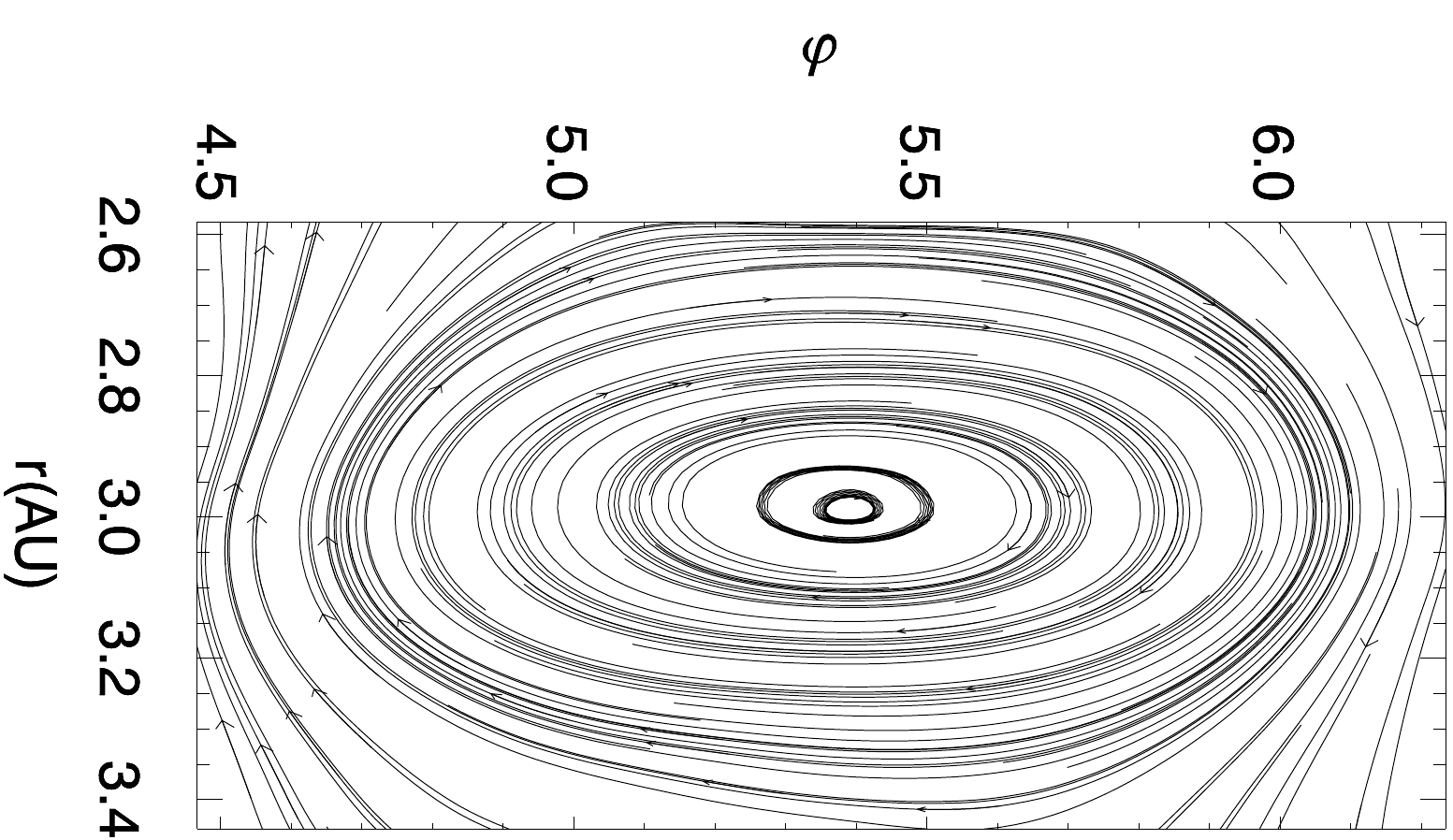}}
      \caption{Vortex streamline at $t=300$ and $t=1000$: the shape of the vortex evolves 
      during the simulation. The radial extent of the vortices is constant and fixed by the 
      radial extent of the density bump, whereas their azimuthal extent increases when the $m$ mode number decreases. On the right figure, the streamlines become elliptic and an aspect ratio can be estimated ($\sim6$ here, but each vortex has a different aspect ratio as can be seen on Fig. \ref{Figevol}).
              }
         \label{Figvortstream}
   \end{figure}

One can notice that the spiral density waves disappear after the growing phase when the vortices survive. 
This implies that Rossby wave instability is not active anymore due to the weakening of the bump by the instability itself and no angular momentum is transferred radially. This is quantified in Fig. \ref{Figaccretion}, where the radial accretion rate is plotted: the bump stops to decrease due to radial accretion.  
When the Rossby vortices are not anymore sustained by the RWI, they should be very efficiently destroyed by differential rotation \citep{TAG01}, but the shape of the streamlines changes to become more similar to closed elliptical streamlines as presented in Fig.~\ref{Figvortstream}.
The vortex streamlines are not directly linked to the shearing sub-Keplerian streamlines, and can survive over more than one rotation time. By the end of the simulation, the vortices have been completely destroyed. This will be discussed in the following section.


      \begin{figure}
   \centering
   \includegraphics[width=8.5cm]{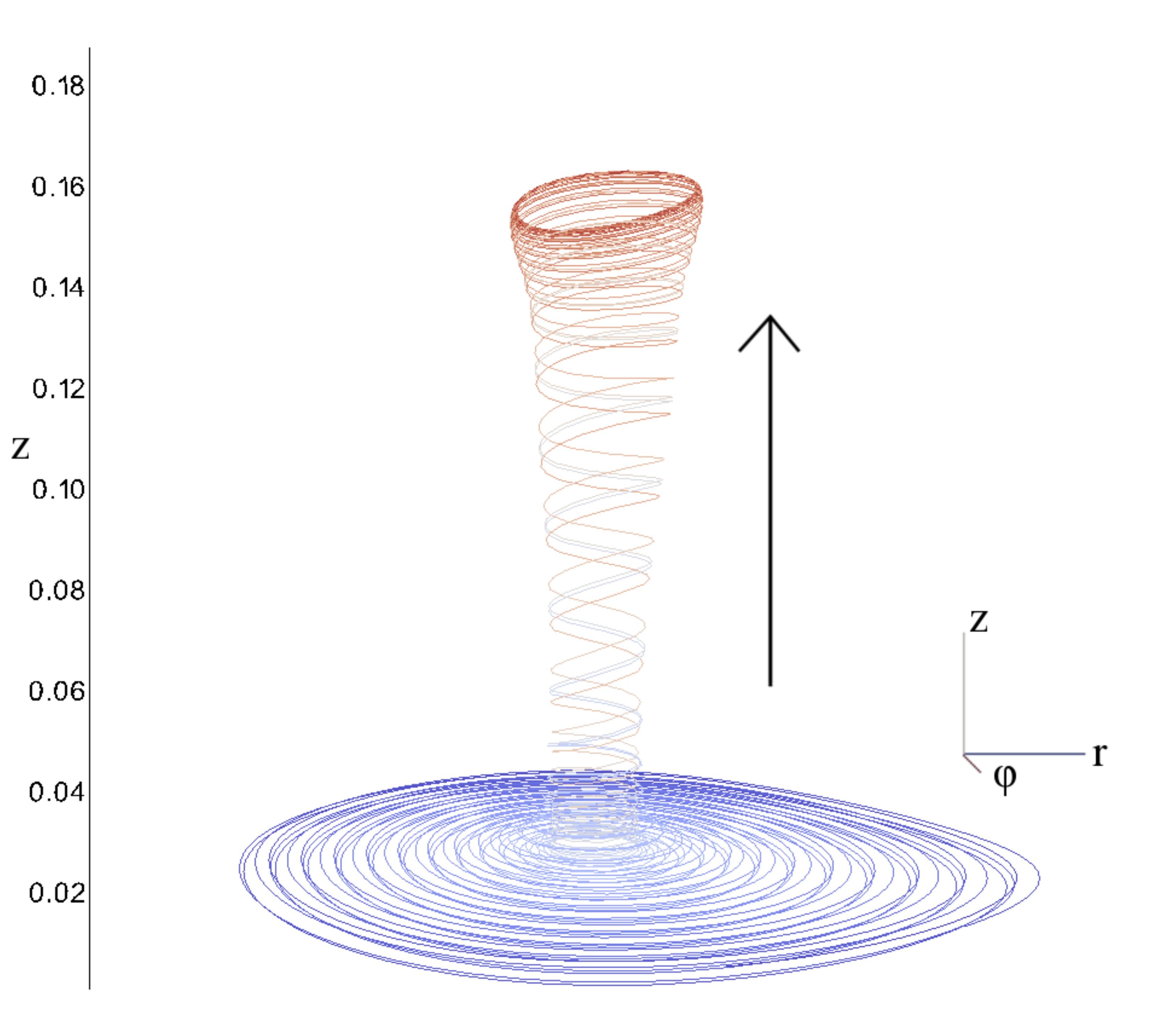}
   \includegraphics[width=8.5cm]{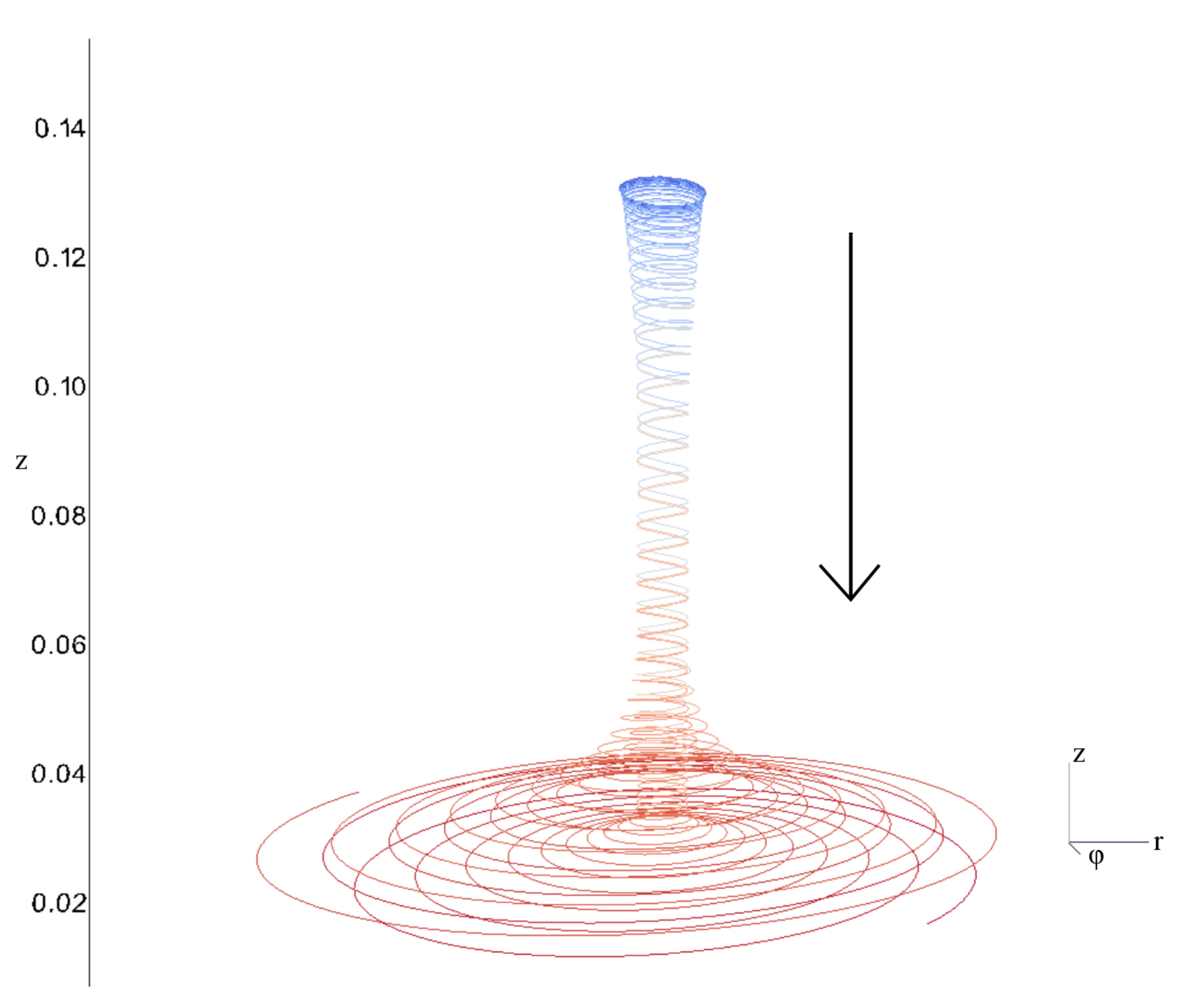}
      \caption{Some streamlines which pass next to the centre of the anticyclone (upper figure) and cyclone (lower figure) at $t=300$ are plotted in 3D. The direction of the flow is given by the colors: the flow moves from the blue part to the red part. The cyclone is a downward flow, contrary to the upward anticyclonic flow. The vortices have a coherent structure over the whole height of the disc. The radial extent of this figure corresponds approximately to half of the one of Fig.\ref{Figvortstream}: only the inner part of the vortices are shown here.
              }
         \label{Figvortstream3D}
   \end{figure}

\section{Discussion}

\subsection{Comparison with previous 3D simulations}

From a numerical point of view, the main difference with the previous simulation of \citet{MEH10} is the use of an AMR grid. This allowed to increase the resolution at the interface between the disk and the corona above it and to avoid numerical difficulties due to dynamics in this low density region. Our current simulation uses an improved limiter and benefits from grid adaptivity, allowing to robustly compute and follow the growth of the instability over longer times. The limit in time is now related to computational resources. 
We tested simulations of the initial growth of the RWI with and without AMR, getting near-identical results.

The simulations presented in this paper show that the most unstable mode of the RWI in the current disc equilibrium has an azimuthal mode number $m=5$. In \citet{MEH10} we stressed the global $m=1$ mode. There, only one mode was present in the seed perturbation, whereas we now choose to use random perturbations that include all modes. For confirmation, we have run the simulation of \citet{MEH10} with random initial perturbations, and the instability was dominated by the $m=5$ mode. The difference with our earlier work is then not due to disparity in initial equilibrium or numerical approach. This result is also coherent with linear calculations of \citet{MYL12}.

The unusual 3D structure of the Rossby vortices with an non-negligible vertical velocity is confirmed by these higher resolution simulations, as can be seen on Fig. \ref{Figvortstream3D}. We have selected some streamlines which pass next to the center of the vortices, they show the `eye' of the vortices and the larger basis. The direction of the flow in the eye of the vortices differs between the cyclones (downward flow) and anticyclones (upward flow). In the outskirts of the vortices, the flows have opposite directions.

\subsection{Vortex migration}


   \begin{figure}
   \centering
   \includegraphics[width=7cm,trim=0 0cm 0 0cm,angle=90]{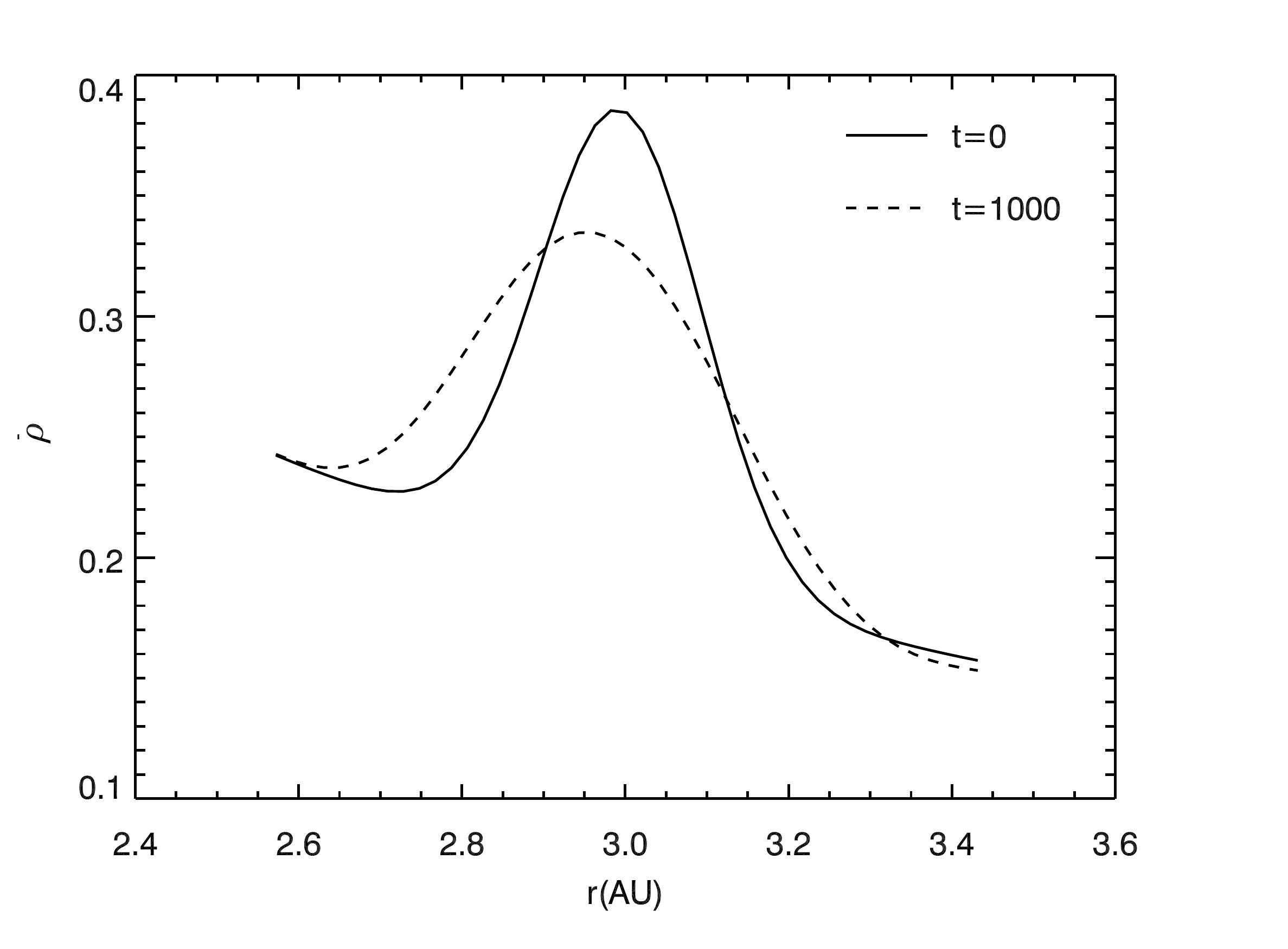}
      \caption{Mean azimuthal density in the mid-plane of the bump region, initially and at 
      $t=1000$. The bump has survived the growth of the RWI.
              }
         \label{Figbump}
   \end{figure}

We don't see any migration of the vortices in our simulation over more than $600$ years. This result is expected, since the growth mechanism of these vortices is localised in the bump region. Vortex migration is oriented in the direction of positive density gradient, and as a consequence the presence of a density maximum at the location of the vortices locks their radial position. One can see on Fig.~\ref{Figbump} that the overdensity is diminished by the Rossby wave instability but it survives the growth of the RWI. The vortices are then indefinitely blocked in this region. 

\subsection{Elliptical instability}\label{elliptic}

The elliptical instability is a purely 3D instability that is due to resonances between the vortex turnover frequency and an inertial wave frequency \citep{KER02}. \citet{LES09} have shown that 3D elliptical vortices are vulnerable to this instability, and can destroy them.
In our simulation, the elliptical instability does not prevent the emergence of vortices by the RWI and their survival over hundreds of years. This was expected as closed velocity streamlines are needed for the turnover frequency to be properly defined and the elliptical instability to grow. Some streamlines of the (non-steady) Rossby vortices can be seen on Fig.~\ref{Figvortstream} (see also Fig.~\ref{Figvortstream3D}). 
After saturation of the RWI, the streamlines tend to close, on the one hand this allows them to survive against shearing, but on the other hand, they become prone to the elliptical instability. 
This may be the reason for the decay of the vortices. The growth of this instability can explain the appearance of smaller scale vertical perturbations (as the ones plotted in Fig. \ref{Figm5vert}) whereas the azimuthal scale is fixed. Indeed contrary to the RWI, the elliptical instability is a local instability without any interaction between the vortices, it can not change their number but affects only their structure. 
The detailed study of the decay of the vortices by the elliptical instability, including elongated vortices or short wavelengths, needs a very high resolution as has been done in \citet{LES09}. 
They studied the effects of the aspect ratio, the ratio of semi-major to semi-minor axis of elliptical streamlines, and showed that elliptical vortices are always unstable. However, the elliptical instability becomes weak for large aspect ratios (i.e. elongated vortices, as is the case in Fig.~\ref{Figvortstream}).  
Therefore, the low azimuthal mode number vortices are expected to survive longer, as observed in our simulations. As pointed out by \citet{GL99-2}, we must caution that the inherently higher numerical dissipation of our finite volume method compared to a spectral method, also induces vortex decay.
The elliptical instability is also present in the vortices formed when the disk is radially buoyant, e.g. subject to a subcritical baroclinic instability, and \citet{LP10} have shown with higher resolution simulations that this secondary instability does not fully destroy the vortices (see also \citealt{LK11}).

\subsection{Outlooks}

We have shown that Rossby vortices can emerge in a 3D protoplanetary disc with an overdensity, and that they can survive for hundreds of years without migrating. When they are not sustained anymore by the RWI, the structure of the vortices will evolve to previously studied vortices in isolation \citep{GL99-2,LES09}. They are then expected to be destroyed and may be prone to the elliptical instability and viscous decay. However, our simulations show that the Rossby vortices can survive for long time scales, if they keep being sustained by the RWI and their unusual structure is preserved. That would be the case if the mechanism that forms the overdensity and launch the RWI, is permanent, such as present in a disc with both active and dead zone regions.

A point to explore is the study of this instability when the active zone is also considered. On the one hand we would expect viscosity to decrease the growth rate of the instability and eventually to destroy the vortices. On the other hand, if a dead zone is included, a density bump arises that will in turn sustain the instability. A full MHD simulation could allow to study jointly the bump formation process and its decay due to the RWI, in combination with magneto-rotational driven accretion.

The results presented here are especially interesting in the context of planetesimal formation. Future work should also study the joint evolution of the gas and dust particles to question the influence of the dust particles on the growth of the RWI and their concentration in Rossby vortices. Previous studies have proposed a bi-dimensional approach \citep{CO10,JAB04,WK05} but a full 3D study, as the one presented here including solid particles, will be necessary to handle both vertical stratification and vortex concentration. 
With this intent, a module of MPI-AMRVAC has been developed and was recently used for circumstellar wind modeling \citep{vMK11}.

On an analytical point of view, the only work we are aware of, that deals with the RWI in 3D is the one by \citet{U10}. 
The vertical structure of the modes should be studied in more detail with dedicated tools as the one presented in \citet{ZL06}. This will be addressed in \citet{MYL12}.

\begin{acknowledgements}
This work was partially supported by the Tournesol program of the PHC (Partenariat Hubert Curien) and by the Swiss National Science Foundation.
\end{acknowledgements}


\end{document}